  \documentstyle[12pt]{article}

  \newcommand{\half}{\frac12}      \newcommand{\threehalf}{\frac32}

     \newcommand{\p}{\vec{p}\,}
     \newcommand{\q}{\vec{q}\,}

     \newcommand{\tdot}{\!\cdot\!}

     \newcommand{\dd}{{\rm d}}       \newcommand{\ee}{{\rm e}}
     \newcommand{\td}{\!{\rm d}}     \newcommand{\ii}{{\rm i}}
     \newcommand{\re}{{\rm Re}\,}

     \newcommand{\bigeqn}{\begin{eqnarray} && \!\!\! \!\!\! \!\!\!}
     \newcommand{\ie}{i.e.}        \newcommand{\Eq}{Eq.\,}
     \newcommand{\cf}{cf.}         \newcommand{\Eqs}{Eqs.\,}

\begin{document}

 \begin{flushright}   Guelph Math. Series 1995-151   \end{flushright}
 \vfill
 {
 \center{\bf{\Large Split Dimensional Regularization for the Coulomb
  Gauge}} \\
 \center{George Leibbrandt\footnotemark\ \\ Department of Mathematics
  and Statistics, University of Guelph, \\ Guelph, Ontario, Canada.
  N1G 2W1 }
 \addtocounter{footnote}{0}
 \footnotetext{E-mail: gleibbra@msnet.mathstat.uoguelph.ca}
 \\
 \center{Jimmy Williams\footnotemark\ \\ Department of Physics,
 University of Guelph \\ Guelph, Ontario, Canada.  N1G 2W1 }
 \addtocounter{footnote}{0}
 \footnotetext{E-mail: williams@physics.uoguelph.ca}

 \vfill \center{December 8, 1995}
        \center{Revised May 24, 1996} \vfill

% \begin{frontmatter}
% \title{Split Dimensional Regularization for the \\ Coulomb Gauge}

% \author[GL]{George Leibbrandt\thanksref{MATH}} and
% \author[JW]{Jimmy Williams\thanksref{PHYS}}

% \address[GL]{Department of Mathematics and Statistics, \\
%  University of Guelph, Guelph, Ontario, Canada.  N1G 2W1 }
% \address[JW]{Department of Physics, \\
%  University of Guelph, Guelph, Ontario, Canada.  N1G 2W1 }

% \thanks[MATH]{E-mail: gleibbra@msnet.mathstat.uoguelph.ca}
% \thanks[PHYS]{E-mail: williams@physics.uoguelph.ca}

\begin{abstract}
\noindent
A new procedure for regularizing Feynman integrals in the noncovariant
Coulomb gauge $ \vec{\nabla } \tdot \vec{A}^a = 0 $ is proposed for
Yang-Mills theory.  The procedure is based on a variant of dimensional
regularization, called {\em split dimensional regularization\/}, which
leads to internally consistent, ambiguity-free integrals.
It is demonstrated that split
dimensional regularization yields a one-loop Yang-Mills self-energy,
$\Pi_{\mu\nu}^{ab}$, that is nontransverse, but local.  Despite the
noncovariant nature of the Coulomb gauge, ghosts are necessary in
order to satisfy the appropriate Ward/BRS identity.  The computed
Coulomb-gauge Feynman integrals are applicable to both Abelian and
non-Abelian gauge models.


 \end{abstract}
 }
 \vfill\eject
 \baselineskip=18pt

% \end{frontmatter}

%%%%%%%%%%%%%%%%%%%%%%%%%%%%%%%%%%%%%%%%%%%%%%%%%%%%%%%%%%%%%%%%%
%
     \section{Introduction}

The quantization of non-Abelian gauge theories in the noncovariant
Coulomb gauge,
  \begin{equation}
  \vec{\nabla } \tdot \vec{A}^a = 0,                      \label{eq:1}
  \end{equation}
has perplexed theorists for decades \cite{1}.  Despite numerous analyses
and ingenious attempts over the past 30 odd years
\cite{2,3,4,5,6,7,8,9,10,11,12,13,14,15,16,17,18,19,20,21}, the Coulomb
gauge has remained an enigma, especially for non-Abelian gauge models
\cite{22,23,24,25,26,27}.  This assessment may come as somewhat of a
surprise in light of the progress made for other noncovariant gauges,
notably the light-cone gauge $n \tdot A^a = 0, n^2 = 0$,
and the temporal gauge $n \tdot A^a = 0, n^2 > 0$, $n_\mu$
being an arbitrary, fixed four-vector \cite{1,28,31,32}.

Our understanding and technical know-how of these axial-type gauges make
it particularly hard to understand why quantization and renormalization
in the Coulomb gauge (also called the {\em radiation gauge\/}) should
have been so elusive \cite{33}.  Could it really be that this gauge is
endowed with characteristics that defy proper definition?  To answer
this question, and in view of the tremendous range of applicability of
the Coulomb gauge in physics generally
\cite{1,34,35,36,37,38,39,40,41,42,43,44,45,46,47,48,49,50,51,52,53,54},
we have decided to take another look at this baffling gauge.

It almost goes without saying that the spurious singularities in the
Coulomb gauge arise specifically from the three-dimensional factor
$(\q^2)^{-1}$ in the gauge propagator $ G_{\mu\nu}^{ab}(q)$,
  \begin{equation}
  G_{\mu\nu}^{ab}(q) = \frac {-\ii \delta^{ab}} {(2\pi)^4 (q^2 + \ii
 \epsilon)} \left[ g_{\mu\nu} - \left( \frac {n^2 q_\mu q_\nu - q\tdot n
 ( q_\mu n_\nu + q_\nu n_\mu ) }{-\q^2} \right) \right],    \label{eq:2}
  \end{equation}
where
  \begin{eqnarray*}
  && \q^2 = q_0^2 - q^2, \qquad \epsilon > 0,      \qquad
     \mu,\nu = 0,1,2,3,  \qquad n_\mu = (1,0,0,0),   \\
  && {\rm diag}(g_{\mu\nu}) = (+1,-1,-1,-1).
  \end{eqnarray*}
Use of this propagator in the gluon self-energy calculation gives rise
to integrals such as
  \begin{equation}
  \int \frac{\dd^4 q\ q_0^2}{[(q+p)^2+\ii \epsilon] \q^2}, \label{eq:4}
  \end{equation}
in which the integration over $q_0$ is UV-divergent because $q_0$ is 
absent from $\q^2$.  Such divergences in the energy integral create
subtle difficulties for the Coulomb gauge which do not occur in other 
popular noncovariant gauges.
Although we could express $(\q^2)^{-1}$ in covariant form, \ie
  \begin{equation}
  \frac1{\q^2} = \frac1{(q \tdot n)^2 - q^2}\,, \label{eq:3}
  \end{equation}
we shall refrain from using this notation, since it deflects attention
from our main goal, which is to find a prescription for $(\q^2)^{-1}$
directly, rather than in the limiting form
  \begin{equation}
  \q^2 = \lim_{\lambda \to 1} [\lambda(q \tdot n)^2 - q^2]. \label{eq:6}
  \end{equation}
Accordingly, the purpose of this article is three-fold:
  \begin{enumerate}
  \item To propose a new procedure, called {\em split dimensional
        regularization\/}, for computing Feynman integrals in the
        noncovariant Coulomb gauge.
  \item To apply the new technique to the one-loop Yang-Mills
        self-energy $\Pi_{\mu\nu}^{ab}$.
  \item To check the appropriate Ward/BRS identity, and hence the
        value of $\Pi_{\mu\nu}^{ab}$.
  \end{enumerate}

Our paper is organized thus.  In Section 2 we summarize the Feynman
rules and state the unintegrated expression for the gluon self-energy
to one-loop order.  The new procedure for evaluating Feynman integrals
is explained in Section 3 and illustrated there by several examples.
The computation of $\Pi_{\mu\nu}^{ab}$ is discussed in Section 4.  In
Section 5, we examine the ghost contributions and verify the appropriate
Ward/BRS identity.  The main features of our calculation are summarized
in Section 6.  Finally, we enumerate in the Appendix some of the
integrals needed for the determination of $\Pi_{\mu\nu}^{ab}$.

%%%%%%%%%%%%%%%%%%%%%%%%%%%%%%%%%%%%%%%%%%%%%%%%%%%%%%%%%%%%%%%%%
%
     \section{Feynman Rules}

The Lagrangian density for pure Yang-Mills theory in the Coulomb gauge,
  \begin{equation}
  \vec{\nabla } \tdot \vec{A}^a = 0, \qquad
  \vec{\nabla } \equiv \left( \frac{\partial }{\partial x}, \frac
     {\partial }{\partial y}, \frac{\partial }{\partial z} \right),
                                                            \label{eq:7}
  \end{equation}
may be written in the form \cite{55}
  \begin{equation}
  {\cal L' = L} - \frac1{2\alpha} \left( {\cal F}_\mu^{ab} A^{b\mu}
     \right)^2, \qquad \alpha \equiv {\rm gauge\ parameter}, \quad
     \alpha \to 0,                                          \label{eq:8}
  \end{equation}
where
  \begin{eqnarray*}
  && {\cal F}_\mu^{ab} \equiv \left( \partial_\mu - \frac{n \tdot
      \partial } {n^2} n_\mu \right) \delta^{ab}, \qquad \mu = 0,1,2,3,
  \\
  && {\cal F}_\mu^{ab} A^{b\mu} = \vec{\nabla } \tdot \vec{A}^a, \qquad
     n_\mu \equiv (n_0, \vec{n}) = (1, \vec{0}), \qquad n^2 = n_0^2 = 1,
  \end{eqnarray*}
and
  \begin{eqnarray*}
  {\cal L} &=& -\frac14 \big( F_{\mu\nu}^a \big)^2 + \big( J_\mu^c +
      \overline\omega^a {\cal F}^{ac}_\mu \big) {\cal D}^{cb\mu}\omega^b
      -\frac12 g f^{abc} K^a \omega^b \omega^c,
  \\
  F^a_{\mu\nu} &=& \partial_\mu A^a_\nu
                 - \partial_\nu A^a_\mu + g f^{abc} A^b_\mu A^c_\nu,
  \\
  {\cal D}^{ab}_\mu &=& \delta^{ab} \partial_\mu + g f^{abc} A^c_\mu.
  \end{eqnarray*}
Here, $g$ is the gauge coupling constant, $f^{abc}$ are group structure
constants, and $A^a_\mu$ denotes a massless gauge field with $a = 1,
\dots, N^2-1$, for SU(N); $\omega^a, \overline\omega^a$ represent ghost,
anti-ghost fields, respectively, while $J^a_\mu$ and $K^a$ are external
sources; the quantities $J^a_\mu, \omega^a, \overline\omega^a$ are
anti-commuting.  The action, $S = \int \dd^4x {\cal L}$, is invariant
under the following Becchi-Rouet-Stora transformations \cite{56}:
  \begin{eqnarray}
    \delta A^a_\mu &=& \lambda {\cal D}^{ab}_\mu \omega^b, \nonumber
    \\
    \delta \omega^a &=& -\frac12 \lambda g f^{abc} \omega^b \omega^c,
                                                            \label{eq:9}
    \\
    \delta \overline\omega^a &=&
            \frac1{\alpha} \lambda {\cal F}^{ab}_\mu A^{b\mu}, \nonumber
  \end{eqnarray}
$\lambda$ being an anti-commuting constant.

The Feynman rules may be summarized as follows.  The {\em gauge boson
propagator\/} in the Coulomb gauge has already been listed in
\Eq (\ref{eq:2}) as \cite{1}
  \begin{equation}
  G_{\mu\nu}^{ab}(q) = \frac {-\ii \delta^{ab}} {(2\pi)^4 (q^2 + \ii
 \epsilon)} \left[ g_{\mu\nu} - \left( \frac {n^2 q_\mu q_\nu - q\tdot n
  ( q_\mu n_\nu + q_\nu n_\mu ) }{-\q^2} \right) \right],  \label{eq:10}
  \end{equation}
$\epsilon > 0$, with components
  \begin{eqnarray} &&
   G^{ab}_{00} = \frac{\ii \delta^{ab}}{(2\pi)^4 \q^2}\,, \qquad
   G^{ab}_{i0} = G^{ab}_{0i} = 0, \qquad i = 1,2,3,           \nonumber
   \\ &&
   G^{ab}_{ij} = \frac{-\ii \delta^{ab}}{(2\pi)^4 (q^2 + \ii \epsilon)}
   \left( -\delta_{ij}+\frac{q_i q_j}{\q^2} \right), \qquad i,j = 1,2,3.
  \end{eqnarray}
The {\em three-gluon vertex\/} \cite{1,30} reads
  \begin{eqnarray}
  V^{abc}_{\mu\nu\rho} (p,q,r) &=& g f^{abc} (2\pi)^4 \delta^4 (p+q+r)
   \nonumber \\ && \cdot \Big[ g_{\mu\nu}(p-q)_\rho
   + g_{\nu\rho}(q-r)_\mu + g_{\rho\mu}(r-p)_\nu \Big],    \label{eq:12}
  \end{eqnarray}
and the {\em scalar ghost propagator\/} (\cf\ \Eq (3.2) of \cite{55}),
  \begin{equation}
  G^{ab}_{\rm ghost} = \frac{\ii \delta^{ab}}{(2\pi)^4 \q^2}\,.
                                                           \label{eq:13}
  \end{equation}

\begin{figure}
\vspace{35mm}
\caption{One-loop gluon self-energy diagram.}
\end{figure}

The {\em unintegrated\/} expression for the one-loop gluon self-energy
(Figure 1), in four-dimensional Minkowski space, is then given by:
  \bigeqn
  \Pi_{\mu\nu}^{ab}(p)\ = \nonumber \\ &&
     \frac {\ii C^{ab}} 2 \int \td^4q \Big[
     g_{\mu\alpha}(q+2p)_\sigma - g_{\alpha\sigma}(2q+p)_\mu
   + g_{\sigma\mu}(q-p)_\alpha \Big] \frac 1 {(q+p)^2 + \ii \epsilon}
  \nonumber \\ &&
    \cdot \left[ g^{\alpha\beta} - \left( \frac
     {n^2 (q+p)^\alpha (q+p)^\beta - (q+p)\tdot n [(q+p)^\alpha n^\beta
     + (q+p)^\beta n^\alpha ] }       {-(\vec{q}+\p)^2} \right) \right]
  \nonumber \\ &&
    \cdot\,\Big[ g_{\beta\nu}(q+2p)_\rho + g_{\nu\rho}(q-p)_\beta
                                        - g_{\rho\beta}(2q+p)_\nu \Big]
  \nonumber \\ &&
    \cdot\,\frac 1 {q^2 + \ii \epsilon}
    \left[ g^{\sigma\rho} -\left( \frac {n^2 q^\sigma q^\rho - q\tdot n
     ( q^\sigma n^\rho + q^\rho n^\sigma ) }{-\q^2} \right) \right],
     \qquad \epsilon > 0,                                  \label{eq:14}
  \end{eqnarray}
where we have defined $f^{acd}f^{bcd} \equiv \delta^{ab}C_{\rm YM}$, and
$C^{ab} \equiv g^2 C_{\rm YM} \delta^{ab}/(4\pi^2)$.
The integral in \Eq (\ref{eq:14}) is divergent; we shall regularize it
with the help of {\em two\/} dimensional parameters, $\omega$
and $\sigma$, for the $\vec{q}$- and $q_0$-integrals, respectively:
  \begin{equation}
  \dd^3\q \to \dd^{2\omega} \q, \qquad
  \dd q_0 \to \dd^{2\sigma} q_0, \qquad p \to P,
  \end{equation}
with the limits $\omega\to \threehalf$ and $\sigma\to \half$ to be taken
after all integrations have been completed.  In this context, the
three-dimensional $\vec{p}$-vector is replaced by the
$2\omega$-dimensional vector $\vec{P}$.
Expansion of the integrand of \Eq (\ref{eq:14}) consequently gives rise
to about 40 noncovariant integrals of the type
  \begin{eqnarray} &&
  \int \frac {\dd^{2\sigma} q_0\,\dd^{2\omega} \q f(q)}
                                     {q^2(\vec{q}+\vec{P})^2}\,,\qquad
  \int \frac {\dd^{2\sigma} q_0\,\dd^{2\omega} \q g(q)}
                              {q^2(q+P)^2(\vec{q}+\vec{P})^2}\,,
  \nonumber \\ &&
  \int \frac {\dd^{2\sigma} q_0\,\dd^{2\omega} \q h(q)}
                     {q^2(q+P)^2\q^2(\vec{q}+\vec{P})^2}\,,\ \dots\ ,
  \end{eqnarray}
where $\ii\epsilon$-terms have been suppressed for clarity.
We describe the methodology for computing these Coulomb-gauge integrals
in Section 3, where the need for two distinct regularizing parameters
will become apparent.

%%%%%%%%%%%%%%%%%%%%%%%%%%%%%%%%%%%%%%%%%%%%%%%%%%%%%%%%%%%%%%%%%
%
     \section{Procedure for Coulomb-gauge integrals}

By a Coulomb-gauge integral we mean any Feynman integral containing one
or more three-dimensional factors such as
  \[
  \frac 1 {\q^2}\,, \qquad
  \frac 1 {(\vec{q}+\vec{P})^2}\,, \qquad {\rm etc.}
  \]
These noncovariant propagators give rise to spurious singularities which
necessarily complicate the integration.  In this section, we propose a
new method for evaluating Coulomb-gauge Feynman integrals.  We shall
illustrate our technique by calculating the Minkowski-space integral
  \begin{equation}
  J \equiv \int \frac {\dd^{2\sigma} q_0\ \dd^{2\omega} \q\ q_0^2}
   {(2\pi)^{2\sigma+2\omega} \,(q^2+\ii\epsilon)(\vec{q}+\vec{P})^2}\,.
                 \label{eq:3.2}
  \end{equation}
Note that if we take $\sigma = \half$, the integration over $q_0$
diverges regardless of the value of $\omega$.  To see the true nature
of this divergence, we may use the identity $q_0^2 = q^2 + \q^2$ to
obtain $J = J_1 + J_2$, where
  \begin{equation}
  J_1 \equiv \int \frac {\dd^{2\sigma} q_0\ \dd^{2\omega} \q}
               {(2\pi)^{2\sigma+2\omega} (\vec{q}+\vec{P})^2}\,, \quad
  J_2 \equiv \int \frac {\dd^{2\sigma} q_0\ \dd^{2\omega} \q
                         \,(\q^2-\ii\epsilon)}{(2\pi)^{2\sigma+2\omega}
                         \,( q^2+\ii\epsilon) (\vec{q}+\vec{P})^2}\,.
  \end{equation}
The divergence as $\sigma \to \half$ occurs only in $J_1$.  Since the
denominator of the integrand of $J_1$ does not involve $q_0$, the
integral is easily factored into
space and time parts, both of which may be shown to vanish using
conventional dimensional regularization:
  \begin{equation}
  J_1 = \int\frac {\dd^{2\sigma} q_0}{(2\pi)^{2\sigma}}
      \int\frac {\dd^{2\omega} \q}{(2\pi)^{2\omega}(\vec{q}+\vec{P})^2}
      = 0, \qquad \re \sigma < 0,\ \re \omega < 1.    \label{eq:J0}
  \end{equation}
In fact, the $q_0$-integral is just a $\delta^{2\sigma}(0)$-integral,
while the $\vec{q}$-integral corresponds to a massless tadpole 
\cite{70,CL}.  We now see the need for two distinct regularizing
parameters:  if either of the limits $\omega\to \threehalf$ or
$\sigma\to \half$ were to be taken before integration, $J_1$ would be
undefined.  

Although a Wick rotation is clearly not needed in the evaluation of
$J_1$, in the general case we may use one provided that we first
integrate out the angular part of $\dd^{2\sigma} q_0$ in the standard
way, \ie,
 \[ \int f(q_0) \dd^{2\sigma} q_0
    \ \to\ \frac {2\sigma\pi^\sigma} {\Gamma (1+\sigma)}
    \int_0^\infty f(q_0) q_0^{2\sigma -1} \,\dd q_0\,.
 \]
The integral over the arc at infinity vanishes when the allowed values
of $\sigma$ are suitably restricted, as in \Eq (\ref{eq:J0}).

For $J_2$, we begin with a Wick rotation to Euclidean space,
  \begin{equation}
  J_2 = \frac {-\ii}{(2\pi)^{2\sigma+2\omega}}
     \int \frac {\dd^{2\sigma} q_4\ \dd^{2\omega} \q
                 \,(\q^2-\ii\epsilon)} { (\vec{q}+\vec{P})^2 q^2}\,,
          \qquad \epsilon\to 0,
  \end{equation}
and then perform the integration in three steps:
\begin{enumerate}
\item It is convenient, although not essential, to use Feynman's
      formula
  \begin{equation}
  \frac 1{AB} = \int_0^1 \td x\,\big[ xA + (1-x)B \big]^{-2},
  \end{equation}
so that
  \begin{equation}
  J_2 = \frac {-\ii}{(2\pi)^{2\sigma+2\omega}} \int_0^1 \td x
     \int \frac {\dd^{2\sigma} q_4\ \dd^{2\omega} \q\ \q^2}
     {[ (1-x)q_4^2 + \q^2 + 2x\q \tdot \vec{P} + x\vec{P}^2]^2 }\,,
  \end{equation}
and then apply exponential parametrization to the denominator:
  \begin{equation}
  J_2 = \frac {-\ii}{(4\pi^2)^{\omega+\sigma}} \int_0^1 \td x
                  \int_0^\infty \td\alpha\,\alpha\ee^{-\alpha G}
                  \int \td^{2\omega}\q\,\q^2     \ee^{-\alpha U}
                  \int \td^{2\sigma} q_4\,       \ee^{-\alpha V},
  \end{equation}
with
 \begin{equation}
   G \equiv x\vec{P}^2,                 \qquad
   U \equiv \q^2 + 2x \q \tdot \vec{P}, \qquad
   V \equiv (1-x) q_4^2.
  \end{equation}
Two points are worth emphasizing:
  \begin{enumerate}
  \item While $V$ in this example is purely quadratic in $q_4$, in
        general $V$ may also contain a term linear in $q_4$.  Hence, it
        is necessary to complete the square in $q_4$ before proceeding
        with the integration.
  \item In contrast to the covariant-gauge case, the coefficient of
        $q_4^2$ (in $V$) differs from that of $\q^2$ (in $U$).
  \end{enumerate}
\item
Since $\q^2$ and $q_4^2$ have unequal coefficients,
we {\em re-scale\/} the $2\sigma$-dimensional $q_4$-vector,
  \begin{equation}
    V = (1-x)q_4^2 = R^2, \qquad
    \dd^{2\sigma}q_4 = (1-x)^{-\sigma}\dd^{2\sigma}\! R,
  \end{equation}
and then use
  \begin{eqnarray}
  \int \td^{2\omega}\q\,\q^2 \ee^{-\alpha U}
   &=& \left( \frac {\pi}{\alpha} \right)^{\omega}
       \left( \frac {\omega}{\alpha} + x^2 \vec{P}^2 \right)
       \exp \big[ \alpha x^2\vec{P}^2 \big],                \nonumber
  \\
  \int \td^{2\sigma}\! R\,\ee^{-\alpha V}
   &=& \left( \frac {\pi}{\alpha} \right)^{\sigma},
  \end{eqnarray}
to obtain
  \begin{eqnarray}
  J_2 &=& \frac {-\ii}{(4\pi^2)^{\omega+\sigma}} \int_0^1 \td x
        \int_0^\infty \td\alpha        \,\alpha\ee^{-\alpha G}
        \int          \td^{2\omega}\q  \,\q^2  \ee^{-\alpha U}
        \int \frac  { \dd^{2\sigma}\! R\,      \ee^{-\alpha V} }
                    { (1-x)^{\sigma} }\,,    \nonumber
  \\
    &=& \frac {-\ii}{(4\pi)^{\omega+\sigma}}     \int_0^1 \dd x
        \int_0^\infty \frac {\dd\alpha\,\big( \omega + \alpha x^2
          \vec{P}^2 \big)}  {(1-x)^{\sigma}\,\alpha^{\omega+\sigma}}
        \,\exp \big[ \alpha (x^2-x) \vec{P}^2 \big]. \label{eq:3.10}
  \end{eqnarray}
\item
The $\alpha$-integration in \Eq (\ref{eq:3.10}) converges if
$\re (\omega+\sigma) < 1$, while the $x$-integration converges if
$\re (\omega+\sigma) > 0$ and $\re \omega > 1$.
Hence, there exists a region in the $\omega\sigma$-plane where the
whole integral is defined.  Performing the integration in this region,
we find that
  \begin{equation}
  J_2 =  \frac {\ii \sigma
        \Gamma(1-\omega-\sigma) \Gamma(\omega-1) \Gamma(\omega+\sigma)}
          {(4\pi)^{\omega+\sigma} \Gamma(2\omega+\sigma-1)}
        \big( \vec{P}^2 \big)^{\omega+\sigma-1}.        \label{eq:3.12}
  \end{equation}
Finally, we analytically continue this result to four-dimensional space
by taking $\omega\to \threehalf$ and $\sigma\to \half$ (in either
order):
  \begin{equation}
    J \,=\, J_1 + J_2 \,=\, -\frac{2\ii}3 \p^2 I_1^*
            \ \ + {\rm\,finite\ terms},               \label{eq:3.13}
  \end{equation}
where $I_1^*$ is defined appropriately by
  \begin{eqnarray}
  I_1^* &\equiv& {\rm divergent\ part\ of\ }
       \int \frac {\dd^{2\omega}\q} {(2\pi)^{2\omega}}
       \int \frac {\dd^{2\sigma} q_4} {(2\pi)^{2\sigma}}
       \frac 1 {q^2 (q+p)^2}\,,           \label{eq:3.14}
  \\
    &=& {\rm divergent\ part\ of\ } \frac {\Gamma(2-\omega-\sigma)
           (p^2)^{\omega+\sigma-2}} {(4\pi)^{\omega+\sigma}}\,,
  \\
    &=& \frac 1 {(4\pi)^2 (2-\omega-\sigma)} \,.
  \end{eqnarray}
Notice that the value of $J$ in \Eq (\ref{eq:3.13}) depends on $\p^2$,
rather than on $p^2$.
\end{enumerate}

The evaluation of $J_1$ in the preceding example hinges decisively on
the use of {\em two\/} complex regulating parameters $\omega$ and
$\sigma$, a drastic departure from conventional dimensional
regularization with its {\em single\/} regulating parameter $\omega$.
The conventional approach was actually applied to the integral
$J$ a couple of years ago by one of the present authors.  Although
the final result looked quite reasonable, its validity was
questioned by J.~C.~Taylor \cite{57}, who noted that the integrals
over the Feynman parameters were ill-defined.

The next example will serve to illustrate the {\em nonlocality\/} of
certain Coul\-omb-gauge integrals.
Consider the integral $I$, containing two covariant propagators, and one
noncovariant propagator:
  \begin{eqnarray}
  I &\equiv& \int^{\rm Mink.} \frac {\dd^{2\sigma}q_0\ \dd^{2\omega}\q}
             {(2\pi)^{2\sigma+2\omega} (q^2+\ii \epsilon)
   [(q+p)^2+\ii \epsilon] (\vec{q}+\p)^2}\,,\qquad \epsilon >0,\nonumber
  \\
    &=& \ii \int^{\rm Eucl.} \frac {\dd^{2\sigma}q_4\ \dd^{2\omega}\q}
             {(2\pi)^{2\sigma+2\omega} q^2 (q+p)^2
           (\vec{q}+\p)^2}\,, \qquad q^2 = q_4^2 + \q^2, \label{eq:3.15}
  \end{eqnarray}
where the same lower case $p$ has been used for convenience for both
the four-vector $p$ and the corresponding
$(2\omega+2\sigma)$-dimensional vector.  Recalling the formula
  \begin{equation}
  \frac 1{ABC} =\int_0^1 \td x \int_0^1 \td z\,z \int_0^\infty \td\alpha
        \,\alpha^2 \exp \big(-\alpha \big[ C+z(B-C)+zx(A-B) \big] \big),
  \end{equation}
we may write \Eq (\ref{eq:3.15}) initially as
  \begin{equation}
  I = \frac {\ii} {(2\pi)^{2\sigma+2\omega}}
      \int_0^1 \td x \int_0^1 \td z\,D,                 \label{eq:3.16}
  \end{equation}
with \begin{eqnarray}
   D &\equiv& z \int_0^\infty \td\alpha\,\alpha^2\ee^{-\alpha G}
      \int \td^{2\omega}\q \, \ee^{-\alpha U}
      \int \td^{2\sigma} q_4\,\ee^{-\alpha V},          \label{eq:3.17}
   \\ G &\equiv& (1-zx)\p^2+z(1-x)p_4^2,                      \nonumber
   \\ U &\equiv& \q^2 + 2(1-zx) \vec{q} \tdot \p,             \nonumber
   \\ V &\equiv& z q_4^2 + 2z(1-x) p_4 q_4
        = z [ q_4 + (1-x) p_4 ]^2 - z (1-x)^2 p_4^2.          \nonumber
  \end{eqnarray}
We then complete the square in $q_4$ (see comment (a) in Step 1), and
execute Step 2 by re-scaling the $q_4$-vector according to
  \begin{equation}
  z [ q_4 + (1-x) p_4 ]^2 = R^2, \qquad
  \dd^{2\sigma} q_4 = z^{-\sigma}\dd^{2\sigma}\! R.
  \end{equation}
Integrating over $\dd^{2\omega}\q$, $\dd^{2\sigma}\! R$, and
then $\dd \alpha$, we readily obtain
  \begin{eqnarray}
  D &=& \frac {\pi^{\omega+\sigma}}{z^{\sigma-1}} \int_0^\infty
        \frac {\dd\alpha}{\alpha^{\omega+\sigma-2}} \,\exp \Big(
       -\alpha zx \big[ (1-x) p_4^2 + (1-zx)\p^2 \big] \Big), \nonumber
  \\
    &=& \frac {\pi^{\omega+\sigma}}{z^{\sigma-1}}
        \frac {\Gamma(3-\omega-\sigma)} {(zx\,p^2)^{3-\omega-\sigma}}
        \left[ 1 - x \left( \frac {p_4^2 + z\p^2}{p^2} \right)
        \right]^{\omega+\sigma-3}.                      \label{eq:3.23}
  \end{eqnarray}
In order to complete the remaining integrations from
\Eq (\ref{eq:3.16}), we first expand the square brackets in
\Eq (\ref{eq:3.23}), and note that only the {\em first\/} term
contributes to the divergent part of $I$ as
$\omega\to \threehalf$ and $\sigma\to \half$ .  Hence,
  \begin{eqnarray}
  I &=& \frac {\ii \Gamma(3-\omega-\sigma) } { (4\pi)^{\omega+\sigma}
               (p^2)^{3-\omega-\sigma} (\omega+\sigma-2) (\omega-1) }
                              \ \ + {\rm\,finite\ terms},  \\
   &=& -\frac{2\ii}{p^2} I^*_1\ \ + {\rm\,finite\ terms},\label{eq:3.25}
  \end{eqnarray}
where $I_1^*$ is defined in \Eq (\ref{eq:3.14}).  Similarly, one may
show that
  \begin{equation}
  \int^{\rm Mink.} \frac {\dd^{2\sigma}q_0\ \dd^{2\omega}\q}
             {(2\pi)^{2\sigma+2\omega}
          (q^2 + \ii \epsilon) [(q+p)^2 + \ii \epsilon] \q^2}
     = -\frac{2\ii}{p^2} I^*_1\ + {\rm\,finite\ terms}.\label{eq:3.26}
  \end{equation}

The appearance of {\em nonlocal\/} Feynman integrals, such as
\Eqs (\ref{eq:3.25}) and (\ref{eq:3.26}), is both necessary and
sufficient for the internal consistency of one-loop integrals in the
Coulomb gauge.  Nor is it entirely unexpected, considering the
noncovariant nature of that gauge.  After all, we have known for some
time that axial gauges likewise lead not only to nonlocal Feynman
integrals, but also to a nonlocal Yang-Mills self-energy \cite{1,29,32}.
We should emphasize that the nonlocality in \Eqs (\ref{eq:3.25}) and
(\ref{eq:3.26}) is {\em not\/} caused by our particular way of
regularizing the integrals, \ie\ by split dimensional regularization,
since the same result is also obtained with conventional dimensional
regularization.

%%%%%%%%%%%%%%%%%%%%%%%%%%%%%%%%%%%%%%%%%%%%%%%%%%%%%%%%%%%%%%%%%
%
     \section{The self-energy $\Pi^{ab}_{\mu\nu}$}

Computations in the Coulomb gauge never seem particularly enjoyable or
uplifting.  Too many trivial things can and do go wrong, and the
compilation of Feynman integrals seems to take forever.  Needless to
say, we were more than relieved to see the various results converge to
manageable form.  For technical reasons, we have chosen to evaluate the
Yang-Mills self-energy $\Pi^{ab}_{\mu\nu}$, \Eq (\ref{eq:14}), in
Euclidean space.  Here is our final result for $\Pi^{ab}_{\mu\nu}(p)$,
written covariantly in Minkowski space:
  \begin{eqnarray}
  \Pi^{ab}_{\mu\nu}(p) &=& \ii C^{ab} \left[
  \frac{11}3 (p^2 g_{\mu\nu} - p_\mu p_\nu)
    -\frac83 (p^2 g_{\mu\nu} - p_\mu p_\nu) \right.\nonumber\\ &&\qquad
  \left. -\,\frac43 \frac{p \tdot n}{n^2} (p_\mu n_\nu + p_\nu n_\mu)
    +\frac83 \frac{p^2 n_\mu n_\nu}{n^2} \right] I_1^*,   \label{eq:4.1}
  \end{eqnarray}
where $n_\mu =(1,0,0,0),\ C^{ab} = g^2 C_{\rm YM} \delta^{ab}/(4\pi^2)$,
and $I_1^*$ is defined in \Eq (\ref{eq:3.14}).  This result for the
Yang-Mills self-energy possesses the following significant features:
  \begin{enumerate}
  \item
$\Pi^{ab}_{\mu\nu}(p)$ is {\em nontransverse\/} in the Coulomb gauge.
  \item
Despite the appearance of {\em nonlocal integrals\/} at intermediate
stages of the computation, $\Pi^{ab}_{\mu\nu}(p)$ is a {\em local\/}
function of the external momentum $p_\mu$.
  \item
Ghosts play an essential role, despite the noncovariant
% ``{\em ghost-free\/}''
nature of the Coulomb gauge.  (See Section 5.)
  \item
Apart from the complex parameters $\sigma$ and $\omega$, defining
{\em split dimensional regularization\/}, no additional parameters are
needed to evaluate $\Pi^{ab}_{\mu\nu}(p)$.
  \item
All one-loop integrals in the Coulomb gauge are ambiguity-free; they are
consistent, at least in the context of split dimensional regularization,
with the values of the following integrals:
  \begin{equation}
  \int\frac {\dd^{2\omega+2\sigma}q\ f(q)} {q^2 \q^2}\,=
  \int\frac {\dd^{2\omega+2\sigma}q\ f(q)} {\q^2 (\vec{q}+\p)^2}\,=
  \int\frac {\dd^{2\omega+2\sigma}q\ f(q)} {(q+p)^2 (\vec{q}+\p)^2}\,=0,
                                                        \label{eq:zeros}
  \end{equation}
where $f(q)$ is any polynomial in the components of $q$.  The latter
integrals are the analogues of tadpole-like integrals which are known
to appear in axial gauges, for example \cite{1}
  \begin{equation}
  \int \frac {\dd^{2\omega}q} {(q\tdot n)^2}\,=\,
  \int \frac {\dd^{2\omega}q} {(q\tdot n) q^2}\,=\,
  \int \frac {\dd^{2\omega}q} {(q\tdot n)((q-p)\tdot n)}\,=0, \qquad
                    {\rm etc.}
  \end{equation}
\end{enumerate}

%%%%%%%%%%%%%%%%%%%%%%%%%%%%%%%%%%%%%%%%%%%%%%%%%%%%%%%%%%%%%%%%%
%
     \section{Verification of the Ward identity}

It has been known for some time \cite{55,56,58,59,60,61,62,63} that
ghosts play a crucial role in the renormalization of non-Abelian
theories, regardless whether the applied gauge is covariant or
noncovariant.  This conclusion holds not only for the noncovariant
gauges of the axial kind, such as the planar gauge and
the light-cone gauge, but also for our Coulomb gauge.  In this section,
we shall examine the role played by ghosts in
obtaining the correct Ward/BRS identity for $\Pi^{ab}_{\mu\nu}(p)$.

Referring to Section 2 for the various definitions of
$S,\ {\cal L,\ L',\ F}_\mu^{ab}$, etc., we recall that the action $S$
satisfies the Becchi-Rouet-Stora identity \cite{56,64,65}
  \begin{eqnarray}
  \sigma S &=& \int \td^4x    \left[ \frac {\delta S}{\delta A^a_\mu(x)}
  \frac {\delta}{\delta J^a_\mu(x)} +\frac {\delta S}{\delta J^a_\mu(x)}
  \frac {\delta}{\delta A^a_\mu(x)} \ + \right.
  \nonumber \\ && \qquad \qquad \left.
      \frac {\delta S}{\delta \omega^a(x)} \frac {\delta}{\delta K^a(x)}
    + \frac {\delta S}{\delta K^a(x)} \frac {\delta}{\delta \omega^a(x)}
    \right] S = 0,
  \end{eqnarray}
and the ghost equation
  \begin{equation}
    \frac {\delta S}{\delta \overline\omega^a (x)} - {\cal F}_\mu^{ab}
    \frac {\delta S}{\delta J_\mu^b (x)} = 0,
  \end{equation}
$\sigma$ being the Slavnov-Taylor operator, $\sigma^2 = 0$.  It is
advantageous to work with the vertex generating functional $\Gamma$
for one-particle-irreducible Green functions with the gauge-fixing
term omitted.  The one-loop divergent parts $D$ of the generating
functional $\Gamma$ must then obey the BRS identity \cite{30,55,61}
  \begin{eqnarray}
    \sigma D &=& \int \td^4x  \left[ \frac {\delta S}{\delta A^a_\mu(x)}
  \frac {\delta}{\delta J^a_\mu(x)} +\frac {\delta S}{\delta J^a_\mu(x)}
  \frac {\delta}{\delta A^a_\mu(x)} \ + \right.
  \nonumber \\ && \qquad \qquad \left.
      \frac {\delta S}{\delta \omega^a(x)} \frac {\delta}{\delta K^a(x)}
    + \frac {\delta S}{\delta K^a(x)} \frac {\delta}{\delta \omega^a(x)}
    \right] D = 0.                                        \label{eq:5.3}
  \end{eqnarray}
Differentiation of \Eq (\ref{eq:5.3}) with respect to $A^b_\nu(y)$ and
$\omega^c(z)$ yields eventually \cite{66}
  \begin{eqnarray}
  \frac {\delta^2(\sigma D)}{\delta\omega^c(z) \delta A^b_\nu(y)}
   &=& \int\td^4x \left[
       \frac {\delta^2 S}{\delta \omega^c(z) \delta J^a_\mu (x)}
       \frac {\delta^2 D}{\delta A^a_\mu (x) \delta A^b_\nu (y)}\ +
  \right. \nonumber \\ && \qquad \left.
       \frac {\delta^2 S}{\delta A^b_\nu (y) \delta A^a_\mu (x)}
       \frac {\delta^2 D}{\delta \omega^c(z) \delta J^a_\mu (x)}
       \right]_{A,J,K,\omega=0} = 0.                      \label{eq:5.4}
  \end{eqnarray}

\begin{figure}
\vspace{25mm}
\caption{Ghost-loop needed for the Ward identity (48).}
\end{figure}

Interpreting the functional derivatives \cite{58}, and
Fourier-transforming to momentum space, we obtain from
\Eq (\ref{eq:5.4}) the following Ward identity in Minkowski space:
  \begin{equation}
  p^\mu \Pi^{ab}_{\mu\nu}(p)
  + (g_{\mu\nu} p^2 - p_\mu p_\nu) H^{ab\mu}(p) = 0,   \label{eq:5.5}
  \end{equation}
or, graphically,
  \begin{equation}
   p^\mu\ \times\ ({\rm Figure}\ 1)
   \ +\ (g_{\mu\nu} p^2 - p_\mu p_\nu)\ \times\ ({\rm Figure}\ 2)\ = 0.
  \end{equation}
It remains to evaluate the ghost contribution $H^{ab\mu}(p)$,
corresponding to Figure 2, and then to check whether the computed
values for $H^{ab\mu}(p)$, together with $\Pi^{ab}_{\mu\nu}(p)$ from
\Eq (\ref{eq:4.1}), respect the Ward/BRS identity (\ref{eq:5.5}).

In order to compute $H^{ab\mu}(p)$, we employ the gluon propagator in
\Eq (\ref{eq:10}), the ghost propagator in \Eq (\ref{eq:13}), the
$J^a$-$A^e$-$\omega^d$ vertex factor $-gf^{aed}$, and the
$A^e$-$\overline\omega^d$-$\omega^c$ vertex factor
$(p_\mu - n\tdot p n_\mu)gf^{dce}$ \cite{55}.  Hence,
  \bigeqn
   H^{ab\mu}(p) \nonumber \\ &
   =& (-\ii^2) C^{ab} \int \frac {\dd^4q\ (p_\beta - n\tdot p n_\beta)}
                                 {(q^2 + \ii \epsilon) (\vec{q}+\p)^2 }
     \left[ g^{\mu\beta} - \left( \frac { q^\mu q^\beta - q\tdot n
     ( q^\mu n^\beta + q^\beta n^\mu ) }{-\q^2} \right) \right],
   \nonumber \\ &
   =& \frac{4i}3 C^{ab} \left(p^\mu - \frac{p\tdot n}{n^2} n^\mu \right)
      I^*_1\,, \qquad \qquad n_\mu = (1,0,0,0),           \label{eq:5.9}
  \end{eqnarray}
which agrees with reference \cite{67}.  We see that the respective
values for $\Pi^{ab}_{\mu\nu}(p)$ in \Eq (\ref{eq:4.1}), and
$H^{ab\mu}(p)$ in \Eq (\ref{eq:5.9}), do indeed satisfy the Ward/BRS
identity (\ref{eq:5.5}).

%%%%%%%%%%%%%%%%%%%%%%%%%%%%%%%%%%%%%%%%%%%%%%%%%%%%%%%%%%%%%%%%%
%
     \section{Conclusion}

In this article we have suggested a new procedure, called {\em split
dimensional regularization\/}, for regularizing Feynman integrals in
the Coulomb gauge \mbox{$\vec{\nabla}\tdot\vec{A}^a$}$=0$.  The
principal feature of this procedure is the use of {\em two\/} complex
parameters, $\omega$ and $\sigma$, which permit us to control more
effectively the respective divergences arising from the $\dd^3\q$- and
$\dd q_4$-integrations.
The method leads to ambiguity-free and internally consistent integrals
which may be either local or {\em nonlocal\/}, and are characterized by
pole terms proportional to $\Gamma(2-\omega-\sigma)$, rather than
$\Gamma(2-\omega)$ (as in conventional dimensional regularization
\cite{68,69,70}).  No additional parameters, apart from $\omega$ and
$\sigma$, are needed to evaluate these integrals.

To test the method of split dimensional regularization at the one-loop
level, we calculated the Yang-Mills self-energy $\Pi^{ab}_{\mu\nu}(p)$.
The latter turned out to be nontransverse, but {\em local\/}, despite
the appearance of {\em nonlocal integrals\/} at intermediate stages of
the computation.  A further check was provided by the Ward/BRS
identity, \Eq (\ref{eq:5.5}), which consists of the self-energy
$\Pi^{ab}_{\mu\nu}(p)$ in \Eq (\ref{eq:4.1}), and the ghost-loop
contribution given in \Eq (\ref{eq:5.9}).  The fact that both
contributions together respect the Ward identity underscores once
again the significance of ghosts, even in the case of noncovariant
gauges such as the Coulomb gauge.

Although the present results seem encouraging, it is too early to
predict whether or not the method of split dimensional regularization
is destined to survive into the 21st century as a viable prescription
for the Coulomb gauge.  Clearly, more calculations are needed,
particularly at two and three loops, before split dimensional
regularization can be placed on a firm mathematical footing, similar
to the successful $n^*_\mu$-prescription for axial gauges.

%   \newpage

%%%%%%%%%%%%%%%%%%%%%%%%%%%%%%%%%%%%%%%%%%%%%%%%%%%%%%%%%%%%%%%%
%
   \vspace{7 mm} \begin{center}  {\bf {\large Acknowledgments}}
                 \end{center}
% \begin{ack}
The first author is deeply indebted to J.~C.~Taylor for his
constructive criticisms concerning the Coulomb gauge, and for numerous
letters on the subject dating back to the winter of 1990.  It gives us
great pleasure to thank A.~K.~Richardson and M.~Staley for discussions
and for performing some preliminary computations, and S.-L.~Nyeo for
showing us how to derive the identity (\ref{eq:5.5}), as well as for
moral support throughout this calculation.  One of us (G.L.) is
grateful to G.~Veneziano, and the staff at CERN, where many of the
integrals were evaluated during the summer of 1995.  The same author
would also like to thank Yu.~L.~Dokshitzer for discussions and for
referring him to reference \cite{36}.  The second author gratefully
acknowledges financial support in the form of an Ontario Graduate
Scholarship, as well as assistance from the Natural Sciences and
Engineering Research Council (NSERC) of Canada.  This research was
supported in part by NSERC of Canada under Grant No.~A8063.
Finally, we would like to thank the referee for his perceptive
comments which have helped us in producing a more readable copy.
% \end{ack}
%%%%%%%%%%%%%%%%%%%%%%%%%%%%%%%%%%%%%%%%%%%%%%%%%%%%%%%%%%%%%%%%%
%
%     \newpage
\ \newline
     \begin{center}     {\bf {\large Appendix}}     \end{center}

Table 1 shows about half of the integrals needed in the evaluation of
$\Pi^{ab}_{\mu\nu}(p)$ and $H^{ab\mu}(p)$.  The others may be obtained
by means of the transformation $p\to -p$, followed by $q\to q + p$,
applied to all components of $p$ and $q$ in $A$, $B$, and
the body of the table.  See also \Eq (\ref{eq:zeros}).

\begin{table}  \caption{Divergent parts of some Coulomb-gauge integrals
  in Euclidean space, as $\omega\to\threehalf$ and $\sigma\to\half$.
  $E_{ijk}\equiv p_i\delta_{jk}+p_j\delta_{ki}+p_k\delta_{ij};\ i,j,k =
  1,2,3$.  All entries are implicitly multiplied by $I_1^*$
  (see \Eq (31)).}

\begin{tabular}{@{}c|c|c|c|c@{}} \hline
 $\stackrel{\displaystyle A}
           {\overbrace{\rule{ 18pt}{0pt}}} $     &  \multicolumn{4}{c}{
 $\stackrel{\displaystyle \int \frac {\dd^{2\omega}\q \dd^{2\sigma}q_4}
             {(2\pi)^{2\omega+2\sigma}} \frac{A}{B} \rule{0pt}{25pt} }
           {\overbrace{\rule{336pt}{0pt}}} $                       } \\
    1  &  2         & $-2/p^2$ & $-2/\p^2$ &   $-4/(\p^2p^2)$ \\
$q_i$  & $-\frac43 p_i$ &   0  &    0      & $2p_i/(\p^2p^2)$ \\
$q_4$  &  0             &   0  &    0      & $2p_4/(\p^2p^2)$ \\ \hline
$q_iq_j$ & $\frac{16}{15} p_ip_j-\frac2{15} \p^2 \delta_{ij}$ &
$\frac13 \delta_{ij}$ & $\frac23 \delta_{ij}$ & $-2p_ip_j/(\p^2p^2)$ \\
$q_iq_4$ &  0             &  0  &   0  & $-2p_ip_4/(\p^2p^2)$ \\
$q_4^2 $ & $-\frac23\p^2$ &  1  & $-2$ & $-2p_4^2 /(\p^2p^2)$ \\ \hline
$q_iq_jq_k$ & --- & $-\frac1{10}E_{ijk} $
                  & $-\frac4{15}E_{ijk} $   & $2p_ip_jp_k/(\p^2p^2)$ \\
$q_iq_jq_4$ & --- & $-\frac16 p_4\delta_{ij}$
                  &      0                  & $2p_ip_jp_4/(\p^2p^2)$ \\
$q_iq_4^2 $ & --- & $-\frac16 p_i$
                  & $ \frac43 p_i$ & $2p_ip_4^2/(\p^2p^2)$ \\ \hline
$\ B\ \Big\{ $ & $q^2(\vec{q}+\p)^2$ & $q^2(q+p)^2\q^2$
  & $q^2(\vec{q}+\p)^2\q^2$ & $q^2(q+p)^2(\vec{q}+\p)^2\q^2$
\\ \multicolumn{5}{c}{\ }
\end{tabular} \end{table}

The integrals in Table 1 were calculated using the efficient technique
described in reference \cite{72}.  Briefly, the most complex $B$ was
first parametrized in accordance with the four-factor analog of
\Eq (\ref{eq:3.16}).  Integration over $\dd^{2\omega}\q$ and
$\dd^{2\sigma}q_4$ was then carried out for the $A=1$ case, and the
result differentiated repeatedly to obtain momentum integrals for the
other eight $A$'s.  Finally, parameter integrations tailored to various
different $B$'s were applied to each of the momentum integrals.

%%%%%%%%%%%%%%%%%%%%%%%%%%%%%%%%%%%%%%%%%%%%%%%%%%%%%%%%%%%%%%%%%
%

     \end{document}